\documentclass[reprint,amsmath,amssymb,pre]{revtex4-2}
\usepackage{graphicx}% Include figure files
\usepackage{dcolumn}% Align table columns on decimal point
\usepackage[caption=false]{subfig}
\usepackage[hidelinks]{hyperref}
\usepackage{cleveref}
\begin{document}

\title{Effect of time-dependent infectiousness on epidemic dynamics}

\author{Nicholas W. Landry}%
\email{nicholas.landry@colorado.edu}
\affiliation{Department of Applied Mathematics, University of Colorado at Boulder, Boulder, Colorado 80309, USA}

\date{October 1st, 2021}

\begin{abstract}
In contrast to the common assumption in epidemic models that the rate of infection between individuals is constant, in reality, an individual's viral load determines their infectiousness. We compare the average and individual reproductive numbers and epidemic dynamics for a model incorporating time-dependent infectiousness and a standard SIR model for both fully-mixed and category-mixed populations. We find that the reproductive number only depends on the total infectious exposure and the largest eigenvalue of the mixing matrix and that these two effects are independent of each other. When we compare our time-dependent mean-field model to the SIR model with equivalent rates, the epidemic peak is advanced and modifying the infection rate function has a strong effect on the time dynamics of the epidemic. We also observe behavior akin to a traveling wave as individuals transition through infectious states.
\end{abstract}

\maketitle

\section{Introduction}

Epidemic modeling has a rich tradition in network science \cite{boguna_epidemic_2002,pastor-satorras_epidemic_2002,newman_spread_2002,pastor-satorras_epidemic_2015} with standard models such as the SIS (Susceptible -- Infected -- Susceptible) and SIR (Susceptible -- Infected -- Removed) models, for which rigorous mathematical theory has been developed. There are also more complex spatio-temporal models that more accurately capture the dynamics of disease spread in the real world \cite{balcan_modeling_2010}. Much interest has been devoted to the accurate prediction of the spread of the SARS-CoV-2 pandemic \cite{arenas_modeling_2020,banerjee_model_2020} and to answering questions such as the efficacy of different prevention measures and the risk factors of different social situations \cite{althouse_superspreading_2020,st-onge_social_2021,larremore_test_2020}. In traditional literature, the SIR model is a canonical example of modeling the spread of disease with total immunity. This model has common extensions such as the SEIR (Susceptible -- Exposed -- Infected -- Recovered) when one wants to incorporate a latent period which captures delays between transmission and infectiousness. With most of these models, however, a key assumption is that an individual's infectivity is constant. However, we know that an individual's infectiousness varies over the duration of the infection, according to their viral load \cite{marks_transmission_2021,he_temporal_2020}. We define a framework to extend the SIR model by dividing the single infectious compartment into $n$ stages as has been considered by Ref.~\cite{lloyd_realistic_2001}, known as the $SI^K R$ model in Ref.~\cite{kiss_mathematics_2017}, and assigning each stage a different infection rate as in Refs.~\cite{ma_generality_2006,hyman_differential_1999}. Other approaches have been considered, such as the message-passing approach \cite{karrer_message_2010, sherborne_mean-field_2018}, mapping an individual's viral load to an infection probability \cite{larremore_test_2020}, and looking at an infection density function \cite{kiss_mathematics_2017, rost_pairwise_2018}. We use this approach to examine fully-mixed populations and theoretical networks constructed from category-based mixing, both static and temporal.

The structure of the paper is as follows. In Section \ref{section:model} we describe a framework for modeling time-dependent infectiousness. In Section \ref{section:r0} we use this model to create theoretical predictions for the reproductive number, apply these predictions to several common cases, and validate our theory with numerical simulations. Lastly, in Section \ref{section:discussion} we discuss the implications of our theory.

\section{\label{section:model} Model}

We propose a general mean-field model to describe the spread of an epidemic including time-dependent infectiousness. In the following, we will refer to this model as the viral load (VL) model.

We consider a population of $N$ nodes. We assume that a node $i$'s intrinsic infectiousness is solely determined by the amount of time it has been infected, $\tau$, and its corresponding viral load at that time, denoted $v_i(\tau)$, although other factors may be involved as well \cite{althouse_superspreading_2020}. Several studies have examined the correspondence between an individual's viral load and their infectiousness \cite{marks_transmission_2021,he_temporal_2020} but for this study, we simply define $\beta_i(\tau)$, the infectious rate function, as the rate at which node $i$ transmits infection having been infected for a duration of time $\tau$. Note that in the case where an infectious threshold exists \cite{mina_rethinking_2020,larremore_test_2020}, we can express the function as $\beta_i(\tau)I_{\tau \in \delta}$, where $\delta=\{\tau \ | \ \beta_i(\tau)\geq \eta\}$ and $\eta$ is the infectious threshold. This infectious rate function can vary in response to many factors such as asymptomatic versus symptomatic infection or severity of symptoms and be considered as being drawn according to some distribution. For much of this study, however, we assume that while $\beta_i(\tau)$ is heterogeneous in time, that every member of the population has the same infectious rate function, i.e., $\beta_i(\tau)=\beta(\tau), \ i=1\dots N$, though we relax this assumption later. We assume that nodes start in the susceptible compartment ($S$) and that an infected individual infected for time $\tau$ infects a susceptible node with rate $\beta(\tau)$. We approximate $\beta(\tau)$ by evaluating it at $n$ discrete times $\tau_j=j\Delta \tau $, where $\Delta \tau$ is fixed and $n\Delta \tau = \tau_R$, the recovery time. Then we divide the infectious compartment, $I$ into $n$ stages, $I_j, \ j=1\dots n$, each with an associated infection rate $\beta_j$, in a similar manner to Refs.~\cite{lloyd_realistic_2001,hyman_differential_1999}. Lastly, nodes that transition through all infection states accumulate in the recovered ($R$) compartment.

\begin{figure}
    \centering
    \includegraphics[width=8.6cm]{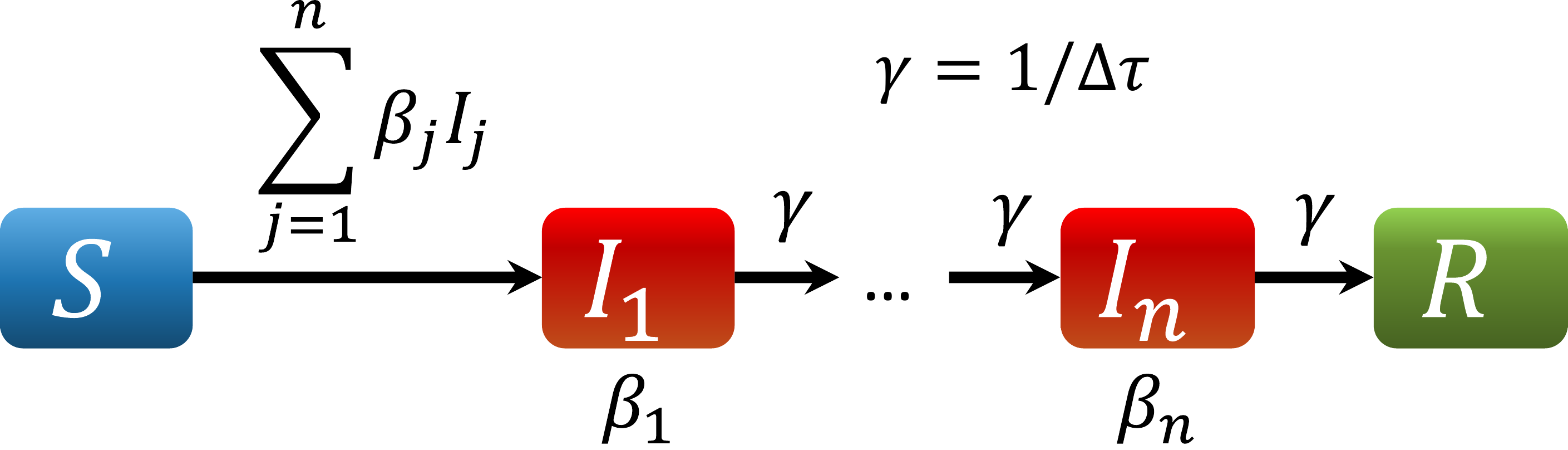}
    \caption{An illustration of the VL model.}
    \label{fig:model}
\end{figure}

We assume that the flow of infected individual between subsequent infectious compartments is deterministic and that upon entering the first infectious stage, an individual passes through all the subsequent stages as shown in Fig.~\ref{fig:model}, meaning that $\gamma_i=1/\Delta \tau$ where $\Delta \tau = \tau_R/n$.

In the following, we define the \textit{m-th moment} of a quantity $q$ as $\langle q^m\rangle=\sum_{i=1}^N q_i^m/N$ when $q$ is a discrete quantity and as $\langle q^m(\tau)\rangle=\int_{0}^{\tau_{R}} [q(\tau)]^m\,d\tau / \tau_{R}$ when $q$ is a continuous function of $\tau$.

There are many studies exploring the effect of more realistic infectious behavior. In Ref.~\cite{lloyd_realistic_2001}, the authors use the $n$-stage $SI^K R$ model with constant infectiousness on a fully-mixed network so that the infectious waiting time is gamma-distributed. In Ref.~\cite{hyman_differential_1999}, the authors explore the $SI^K R$ model with variable infectiousness for fully-mixed networks. For both of these models, the authors allow healing and recovery to occur at every infectious stage. In Ref.~\cite{sherborne_dynamics_2015}, the authors explore the $SI^K R$ link-closure model with a constant infection rate and solely consider static networks. They simulate their model numerically on homogeneous and Erd\"os-R\'enyi networks. In Ref.~\cite{karrer_message_2010}, the authors consider a message-passing approach to model time-dependent infectiousness and simulate their results on a static network. In Ref.~\cite{sherborne_mean-field_2018}, the authors present a non-Markovian edge-based compartment model, prove its equivalence to the message-passing model, and describe how other models compare to the message-passing approach. In Refs. \cite{lloyd_realistic_2001,hyman_differential_1999} the authors solely consider the fully-mixed case and in Refs.~\cite{sherborne_dynamics_2015,sherborne_mean-field_2018,karrer_message_2010} the authors solely consider static networks. In contrast, our approach encompasses fully-mixed, static, and temporal networks. In Refs.~\cite{lloyd_realistic_2001,sherborne_dynamics_2015}, though the authors consider an $SI^K R$ model, they specify that the infectious rate is constant in contrast to our model where we allow the rate to vary over time. In addition, in Refs.~\cite{lloyd_realistic_2001,hyman_differential_1999,sherborne_dynamics_2015,sherborne_dynamics_2015}, they assume Markovian transitions between infectious states in contrast to our approach which enforces deterministic transitions between infectious states (as in Ref.~\cite{larremore_test_2020}).

\section{\label{section:r0} Derivation of the population reproductive number}

We derive the reproductive number for the viral load model described above that has been cast as a system of mean-field ODEs. First, we derive the reproductive number for a fully-mixed model and second, we derive the reproductive number for an arbitrary category-mixed population. We comment on the continuum limit for both cases and derive specific closed-form solutions for the reproductive number for a configuration model static network, and an activity model temporal network.

\subsection{\label{section:r0fullymixed} Fully-mixed population}

Consider a fully-mixed population of $N$ individuals and an infectious rate function, $\beta(\tau)$. In our formalism, we denote the fraction of the population in the susceptible, $j$th infectious stage, and the recovered stage as $S$, $I_j,\,j=1\dots n$, and $R$ respectively and note that $S+\sum_{j=1}^n I_j + R=1$ by conservation. Assuming that an individual's infection status is independent of the infection status of its neighbors, as done in Ref.~\cite{hyman_differential_1999}, we can write the following system of mean-field equations as

\begin{subequations}
\begin{align}
\frac{dS}{dt}&=-S\sum_{j=1}^n \beta_j I_j,\\
\frac{dI_1}{dt}&=-\frac{I_1}{\Delta \tau}+S\sum_{j=1}^n \beta_j I_j,\\
\frac{dI_j}{dt}&=\frac{I_{j-1}-I_j}{\Delta \tau}, \ j=2\dots n,\label{eqn:fullymixed_transport}\\
\frac{dR}{dt}&=\frac{I_{n}}{\Delta \tau}.
\end{align}
\end{subequations}

By construction, an infected node will always transition through all the infectious states until it reaches the recovered state. However, we are not interested in whether infected nodes transition through all the states, but rather whether susceptible nodes become infected. In Ref.~\cite{diekmann_construction_2010}, the authors introduce the notion of a \textit{next generation matrix} (NGM) which decomposes the linearized system into infectious transmissions, $T$, and non-infectious transitions, $\Sigma$, where transmissions move susceptible nodes to infected compartments and transitions move infected nodes to other infectious states. As done in Ref.~\cite{diekmann_construction_2010}, we exclude the susceptible and recovered states. The linearized system can be written as

\begin{equation*}
   {\bf I}'=\frac{1}{\Delta\tau}\begin{pmatrix}-1 +\beta_1\Delta\tau & \beta_2 \Delta\tau & \dots & \dots & \beta_n\Delta \tau \\ 1 & -1 & 0 & \dots & 0\\ 0 & 1 & -1 & \ddots & \vdots\\ \vdots & \ddots & \ddots & \ddots & 0\\ 0 & \dots & 0 & 1 & -1 \end{pmatrix}{\bf I},
\end{equation*}
where ${\bf I}=(I_1,\dots, I_n)^T$. We split the matrix into transmissions and transitions and according to Ref.~\cite{diekmann_construction_2010}, the reproductive number $R_0$ is given by $\rho(-T\Sigma^{-1})$ which for the fully mixed case evaluates to
\begin{equation}
    R_0=\sum_{i=1}^n \beta_i\Delta \tau,
\end{equation}
which matches the value found in Ref.~\cite{hyman_differential_1999}.

This result indicates that any infectious rate function that has the same total infectiousness or \textit{exposure} yields the same reproductive number, regardless of the particular function. This, however, does not hold for the time scale on which the epidemic spreads as we will see later.

\subsection{\label{section:r0categorymixed} Discrete category-mixed population}

Now we consider a population with $N$ individuals each of which belong to a category $c_i, i=1 \dots n_c$. These mixing categories can encode many different characteristics such as degree-based mixing \cite{miller_edge-based_2012}, age-mixing \cite{mistry_inferring_2021}, spatial meta-population mixing \cite{balcan_modeling_2010}, mixing due to travel and many other types of mixing.

We denote the probability that sub-populations $c_i$ and $c_j$ interact with each other as $p(c_i, c_j)$ and the probability that a node belongs to category $i$ as $p(c_i)$. We discretize the infectious states not only by the progression of the infection, but by the category to which that individual belongs as well. This model has $(n+2)n_c$ states: $n_c$ susceptible states, $S^{c_1},\dots,S^{c_{n_c}}$; $n n_c$ susceptible states, $I_1^{c_1},\dots,I_1^{c_{n_c}},\dots, I_n^{c_1},\dots,I_n^{c_{n_c}}$; and $n_c$ recovered states, $R^1,\dots,R^{n_c}$. Then the mean-field model becomes for each category $c$

\begin{subequations}
\begin{align}
\frac{dS^{c}}{dt}&=-S^c\sum_{i=1}^{n_c}\sum_{j=1}^n p(c,c_i)p(c_i) \beta_j I_j^{c_i},\\
\frac{dI_1^{c}}{dt}&=-\frac{I_1^{c}}{\Delta \tau}+S^c\sum_{i=1}^{n_c}\sum_{j=1}^n p(c,c_i)p(c_i) \beta_j I_j^{c_i},\\
\frac{dI_j^{c}}{dt}&=\frac{I_{j-1}^c-I_j^c}{\Delta \tau}, \ j=2\dots n,\label{eqn:categories_transport}\\
\frac{dR^{c}}{dt}&=\frac{I_n^c}{\Delta \tau}.
\end{align}
\end{subequations}

The linearized ODE is the following block-matrix system of equations:
\begin{equation*}
{\bf I}' = \frac{1}{\Delta\tau}\begin{pmatrix} -I +\beta_1\Delta \tau P & \beta_2 \Delta \tau P & \dots & \dots & \beta_n \Delta \tau P\\ I & -I & 0 & \dots & 0\\ 0 & I & -I & \ddots & \vdots\\ \vdots & \ddots & \ddots & \ddots & 0\\ 0 & \dots & 0 & I & -I \tau \end{pmatrix}{\bf I},
\end{equation*}
where
\begin{equation*}
P = \begin{pmatrix} p(c_1,c_1)p(c_1) & \dots & p(c_1,c_{n_c})p(c_{n_c}) \\ \vdots & \ddots & \vdots \\ p(c_{n_c},c_1)p(c_1) & \dots & \beta_i p(c_{n_c},c_{n_c})p(c_{n_c}) \end{pmatrix},
\end{equation*}
${\bf I}=(I_1^{c_1}, \dots, I_1^{c_C{n_c}}, \dots, I_n^{c_1}, \dots, I_n^{c_{n_c}})^T$, and $I$ is the identity matrix.

Splitting the matrix into transmissions and transitions, the next-generation matrix is
\begin{equation}
-T\Sigma^{-1}=\begin{pmatrix} P\sum_{i=1}^n \beta_i\Delta \tau & P\sum_{i=2}^n \beta_i \Delta \tau & \dots & P \beta_n \Delta \tau\\ 0 & \dots & \dots & 0\\ \vdots & \ddots & \ddots & \vdots \\ 0 & \dots  & \dots & 0 \end{pmatrix}.
\end{equation}

Then, the reproductive number evaluates to
\begin{equation}
    R_0 = \rho(P)\sum_{i=1}^n \beta_i \Delta \tau,
\end{equation}
which indicates that the epidemic threshold depends both on the infectious exposure and the matrix of mixing probabilities and that these two quantities are independent.

\subsection{\label{section:continuum} The continuum limit} For each case described prior, it is natural to want to take the limit as the number of infectious compartments approaches infinity and $\Delta \tau \to 0$. For the fully-mixed case, the reproductive number becomes
\begin{equation}
    R_0=\int_0^{\tau_R}\beta(\tau)d\tau,
\end{equation}
and similarly, for category-based mixing, it is
\begin{equation}
    R_0=\rho(P)\int_0^{\tau_R}\beta(\tau)d\tau.
\end{equation}

Alternatively, we can treat $\tau$ as a continuous quantity and track the infectiousness, $I(t,\tau)$, as a function of the overall time and how long an individual has been infected. When $\tau$ is continuous, $\Delta\tau \to 0$ and the finite difference $(I_{j-1}-I_j)/\Delta \tau$ in Eqns. \eqref{eqn:fullymixed_transport} and \eqref{eqn:categories_transport} becomes a derivative with respect to $\tau$. With these assumptions, our ODE model can be expressed as the transport equation with boundary conditions handling the infection and recovery. For the fully-mixed case, this is
\begin{subequations}
\begin{align}
\frac{\partial I(t,\tau)}{\partial t} &= -\frac{\partial I(t, \tau)}{\partial \tau},\\
I(t, 0) &= S \int_0^{\tau_R} \beta(\tau) I(t, \tau) d\tau,\\
S &= 1 - \int_0^{\tau_R} I(t, \tau) d\tau - \int_0^t \frac{\partial I(t,\tau)}{\partial \tau}\bigg{|}_{\tau=\tau_R} dt,\\
I(t, \tau_R) &= 0.
\end{align}
\end{subequations}

\begin{figure}
    \centering
    \includegraphics[width=8.6cm]{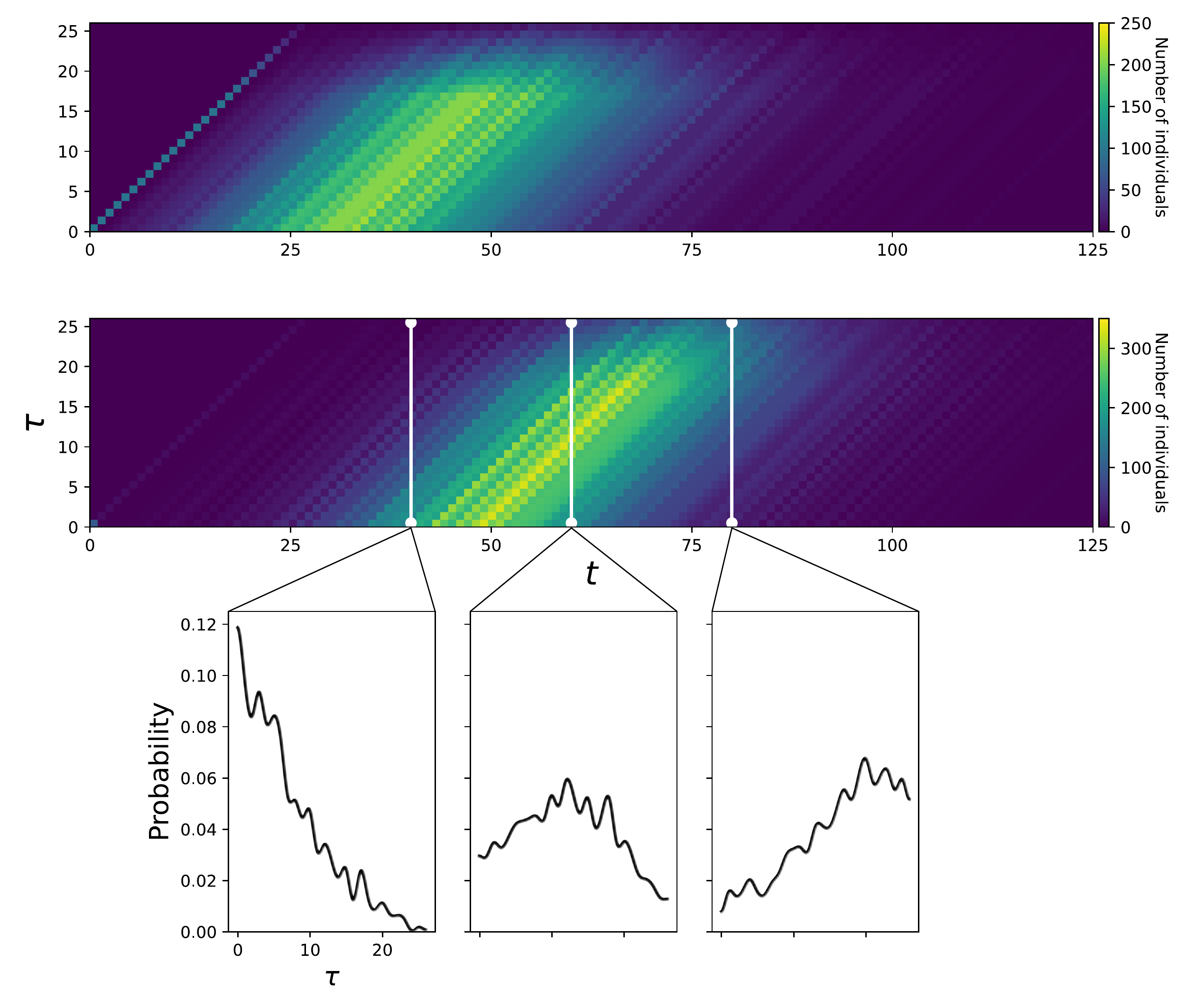}
    \caption{The number of individuals infected for duration $\tau$ at time $t$ for the configuration model (top) and the activity model (middle). The line plots (bottom) denote the probability distribution of $\tau$ at times 40, 60, and 80, which correspond to normalized vertical cross-sections of $I(t,\tau)$.}
    \label{fig:heatmap}
\end{figure}

The transport equation admits traveling wave solutions and this perspective lends physical interpretation to our model; an infected individual is transported through the infectious stages and the boundaries merely introduce new individuals into this transport process and remove recovered individuals at the other boundary. We can see this behavior in Fig.~\ref{fig:heatmap} for both static and temporal networks.

\begin{figure}
    \centering
    \includegraphics[width=8.6cm]{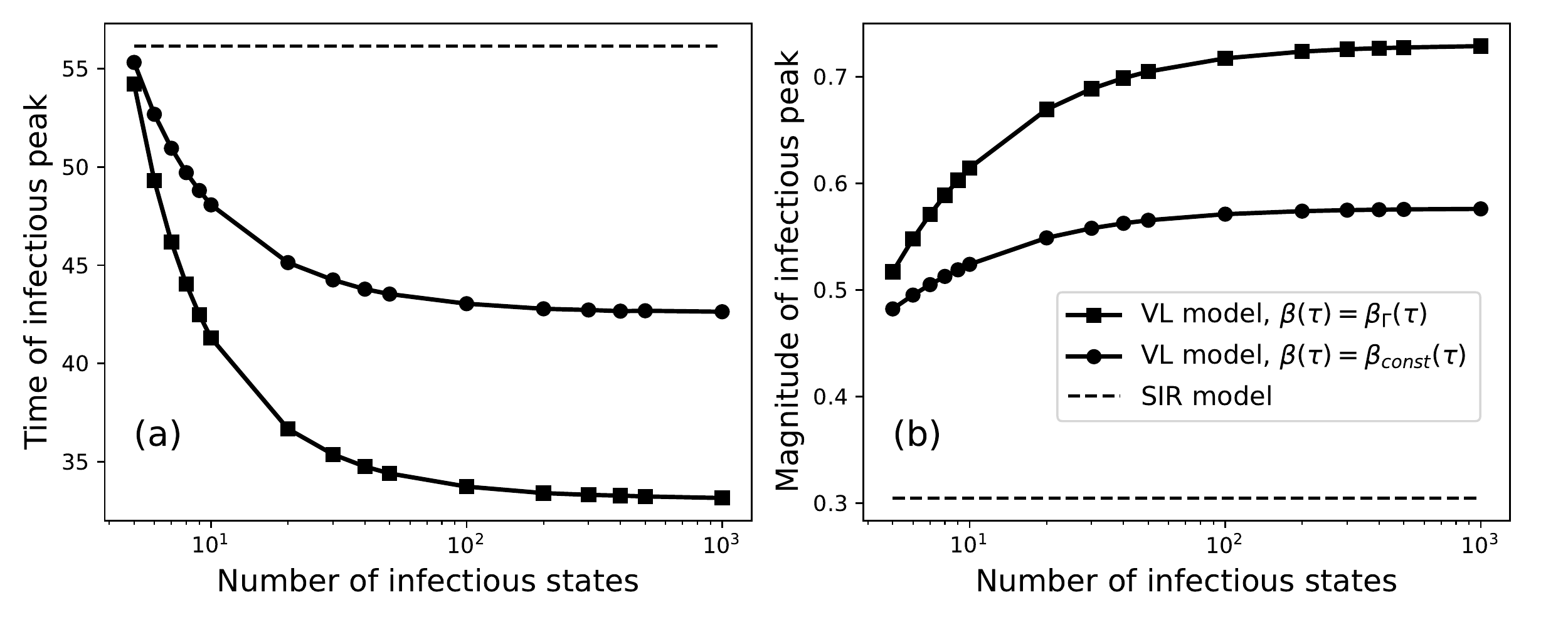}
    \caption{A plot showing how the number of infectious states affects (a) the time at which the infectious peak occurs and (b) the magnitude of the infectious peak for the viral load model in the fully-mixed case. We use two different infectious rate functions described in Section \ref{section:numerical_experiments} and show the constant value of the SIR model (1 infectious stage) as a reference. For every data point, $R_0 = 3$.}
    \label{fig:mf_peak_difference}
\end{figure}

Because our approach approximates the infectious rate function with discrete infectious compartments, we perform numerical experiments to analyze the number of states at which we can expect the mean-field ODE model to reasonably approximate the continuous rate function. For a small number of states, the discretized values of the infectious rate function fluctuate, leading to non-monotone and non-smooth trends, so we only look at the viral load model with greater than 4 infectious states. As the number of infectious states is increased, the epidemic dynamics converge to that of the continuous VL model with a continuous infectious rate function. From Fig.~\ref{fig:mf_peak_difference}, approximately 100 infectious states are necessary to capture key features of the epidemic response.

\subsection{\label{section:r0examples} Examples} In the following, we apply our category-mixing framework to two cases, a static degree-based configuration model and a temporal activity-based model.

\subsubsection{\label{r0configmodel} Configuration model}
Consider a network of size $N$ with a degree sequence ${\bf k} = (k_1,\dots,k_N)^T$ and nodes connected by links at random, which specifies the configuration model (described more in Ref.~\cite{fosdick_configuring_2018}). Networks generated with the configuration model may have a non-negligible number of self-loops and multi-edges in the infinite size limit \cite{catanzaro_generation_2005}, leading to correlated simple networks. In this study, however, we consider a bounded degree distribution and so we can assume the configuration model to be uncorrelated for large enough $N$. For the standard SIR model on a configuration model network, the reproductive number is $R_0 = \beta\langle k^2\rangle/(\gamma\langle k\rangle)$ \cite{boguna_epidemic_2002}. We assume that a node's degree completely specifies its dynamic behavior, which ignores effects from a node's other characteristics. From the degree sequence ${\bf k}$, we can compute the discrete probability distribution $p(k) = N(k)/N$, where $N(k)$ is the number of nodes in the degree sequence that have degree $k$, and the list of unique degrees in the degree sequence, ${\bf k}_u$. From our general formalism in Section \ref{section:r0categorymixed}, the degree mixing matrix is
\begin{equation}
P = \frac{1}{\langle k\rangle}({\bf k}_u{\bf p})^T {\bf k}_u.
\end{equation}
where ${\bf k}_u{\bf p} = (k_1 p(1), \dots, k_{max} p(k_{max}))^T$ and ${\bf k}_u=(k_1,\dots,k_{max})^T$. The largest eigenvalue of this matrix is $\langle k^2\rangle/\langle k\rangle$ and so the reproductive number is

\begin{equation}
    R_0 = \frac{\langle k^2\rangle}{\langle k\rangle}\int_0^{\tau_R}\beta(\tau)d\tau.
\end{equation}

Setting $\gamma = 1/\tau_R$ and $\beta = \langle \beta(\tau)\rangle = \int_0^{\tau_R}\beta(\tau)d\tau/\tau_R$ for the SIR model yields the reproductive numbers derived in Ref.~\cite{boguna_epidemic_2002}.

\subsubsection{\label{section:r0activitymodel} Activity model}

Our category-based framework applies not only to static contact structures, but to temporal networks as well. We consider the \textit{activity model} first presented in Ref.~\cite{perra_activity_2012}. Given a temporal network of size $N$, suppose that each node $i$ has an \textit{activity rate} $a_i$, which denotes the probability per unit time that the node is active. At each discrete time, each node is either active or idle, and each active node forms $m$ connections with other nodes, active or inactive. Unlike degrees which are discrete for an unweighted network, these activity rates are continuous, and to use our category-based mixing framework, we assume that we can bin these rates into discrete categories, $a_i, i = 1\dots n_a$ and later take the continuum limit as before. We denote the probability that a node has an activity rate $a_i$ as $p(a_i)$. Then the probability that nodes with activity rates $a_i$ and $a_j$ are connected at any given time is $(a_i+a_j)\frac{m}{N}$ and the time-averaged mixing matrix is
\begin{equation*}
    P_{ij}=\frac{m(a_i+a_j)}{N}p(a_j),
\end{equation*}
which can be written $P = \mathbf{1b}^T + \mathbf{cp}^T$ where $\mathbf{b}=(m\,a_1\,p(a_1),\dots,m\,a_{n_a}\,p(a_{n_a}))^T$, $\mathbf{c}=(m\,a_1,\dots,m a_{n_a})^T$, and ${\bf p}=(p(a_1),\dots, p(a_{n_a}))^T$. Observing that this is a rank-2 matrix, the analytical solution for the Perron-Frobenius eigenvalue is $(m\langle a\rangle + m \sqrt{\langle a^2\rangle})$ and
\begin{equation}
    R_0=(m\langle a\rangle + m \sqrt{\langle a^2\rangle})\int_0^{\tau_R} \beta(\tau) d\tau.
\end{equation}

In Ref.~\cite{perra_activity_2012}, they derive the epidemic threshold for the activity model as $\beta/\gamma=2\langle a\rangle/(\langle a\rangle + \sqrt{\langle a^2\rangle})$. As before, setting $\gamma=1/\tau_R$ and $\beta = \langle k \rangle\langle \beta(\tau)\rangle = 2m\langle a \rangle\langle \beta(\tau)\rangle$ yields the same result.

\subsection{\label{section:individual_variation} Individual variation in the infectious rate function}

In Ref.~\cite{gou_how_2017}, the authors consider heterogeneous susceptibility and recovery rate for the SIR model. Similarly, we now relax the assumption that the infectious rate function is the same for every individual. We extend our results in Section~\ref{section:r0categorymixed} for a distribution of infectious rate functions over the population. In our analysis, we assume that the particular infectious rate function is distributed independently of any other nodal characteristic such as its degree. We denote $p_b(b)$ as the fraction of the population with an infectious rate function of $\beta_b(\tau)$ and an associated recovery time of $\tau_{R_b}$, where the number of unique infectious rate functions is $n_b$. We enforce that the number of infectious states regardless of recovery time is $n$ so the time between infectious compartments is $n\Delta \tau_b = \tau_{R_b}$. We define the discretized values $\beta_i(\tau_j) = \beta_i(j \Delta \tau_{i})$ as $\beta_j^i$ and denote the $j$th infectious stage with infectious rate function $\beta_b(\tau)$ and category $c$ as $I_j^{b,c}$. Then the mean-field equations become
\begin{subequations}
\begin{align}
\frac{dS^{b,c}}{dt}&=-S^{b,c}\sum_{i=1}^{n_b} \sum_{j=1}^{n_c} \sum_{k=1}^{n} p_b(b_i) \beta_k^{b_i}  p(c,c_j)p(c_j)I_k^{b_i,c_j},\\
\frac{dI_1^{b,c}}{dt}&=-\frac{I_1^{b,c}}{\Delta \tau}\nonumber\\
&+ S^{b,c}\sum_{i=1}^{n_b} \sum_{j=1}^{n_c} \sum_{k=1}^{n} p_b(b_i) \beta_k^{b_i}  p(c,c_j)p(c_j)I_k^{b_i,c_j},\\
\frac{dI_j^{b,c}}{dt}&=\frac{I_{j-1}^{b,c}-I_j^{b,c}}{\Delta \tau}, \ j=2\dots n,\\
\frac{dR^{b,c}}{dt}&=\frac{I_n^{b,c}}{\Delta \tau}.
\end{align}
\end{subequations}

Linearizing these equations, we obtain ${\bf I}' = A{\bf I}$, where $A = \Sigma + T$. $\Sigma$ and $T$ are each $n\times n$ block matrices of size $n_c n_b \times n_c n_b$ with blocks of size $n_c \times n_c$.
\begin{align*}
\Sigma_{i,j} = \begin{cases} \text{diag}(I/\Delta \tau_1,\dots, I\Delta \tau_{n_b}), & i=j\\ \text{diag}(-I/\Delta \tau_1,\dots, -I\Delta \tau_{n_b}),& i = j + 1 \end{cases}
\end{align*}
and
\begin{align*}
T_{i,j} = \begin{cases} \begin{pmatrix} p_b(b_1) \beta_j^1 P & \dots & p(b_{n_c}) \beta_j^{n_b} P\\ \vdots & \ddots & \vdots \\ p_b(b_1) \beta_j^1 P & \dots & p(b_{n_b}) \beta_j^{n_b} P \end{pmatrix},& i = 1\\[0.3in] {\bf 0},& \text{else}.\end{cases}
\end{align*}
Then, the reproductive number (the maximal eigenvalue of $-T\Sigma^{-1}$) is
\begin{align*}
R_0 = \rho(P)\sum_b p_b(b)\sum_{j=1}^{n}\beta_j^b\Delta \tau_{b}.
\end{align*}
As $n \to \infty$, $\Delta \tau_b \to 0$ for every $b$ and we obtain
\begin{align}
R_0 = \rho(P)\sum_b p_b(b)\int_0^{\tau_{R_{b}}}\beta_b(\tau)d \tau,
\end{align}
which is the value obtained for the category-mixed case with the key difference that the \textit{exposure} is now the average exposure with respect to the distribution of infectious rate functions.

\subsection{\label{section:numerical_experiments} Numerical experiments} We compare the time dynamics of the SIR model with that of the VL model with different infectious rate functions. For the following figures, we fixed $N=10^4$, $R_0=3$, $\tau_R = 21$ days, and $\arg\max_{\tau}\beta(\tau)=4$ days, unless otherwise noted. We considered the configuration model with a power-law degree distribution $p(k)\propto k^{-3}$ on $[10,1000]$ and the activity model with activity rates $p(a)\propto k^{-3}$ on $[0.01, 1]$, $m = 10$, and $\Delta t = 1$. We used the following contagion models: the VL model with $\beta_{\Gamma}(\tau)\propto \tau \exp(\tau/4)$ as in Ref.~\cite{he_temporal_2020}, the VL model with a constant-valued infectious rate function, $\beta_{const}(\tau) = \langle \beta_{\Gamma}(\tau)\rangle$, and the SIR model with a single infectious rate of $\beta = \langle \beta_{\Gamma}(\tau) \rangle$ for the configuration model and $\beta = 2m\langle a\rangle \langle \beta_{\Gamma}(\tau) \rangle$ for the activity model. These relations were chosen such that the reproductive numbers are the same for each infection model.

\begin{figure}
    \centering
    \includegraphics[width=8.6cm]{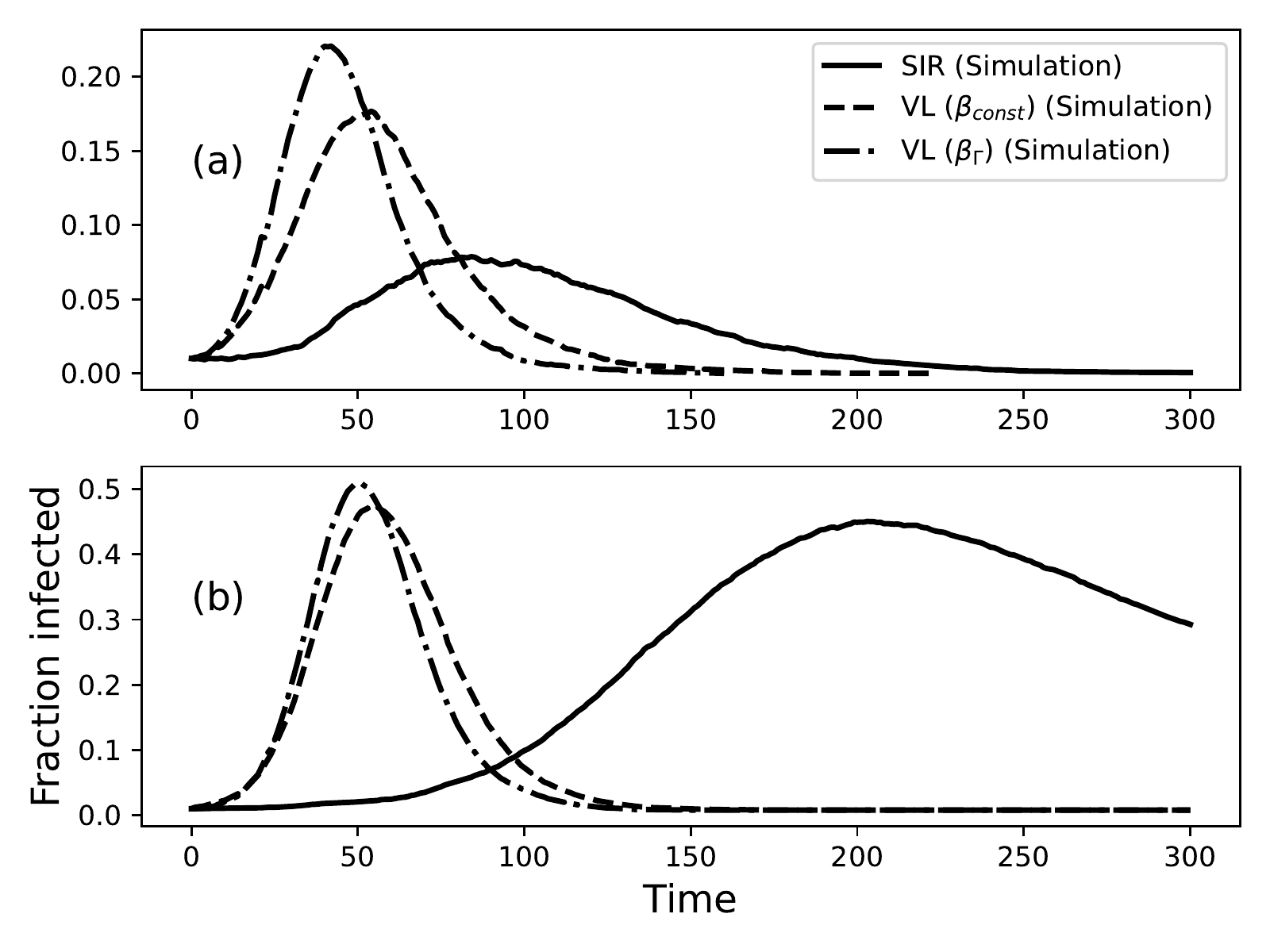}
    \caption{Time response of the fraction of infected individuals for different contagion models for (a) the configuration model and (b) the activity model. For both (a) and (b), the dash-dot, dashed, and solid lines indicate the VL model with $\beta_{\Gamma}(\tau)\propto \tau \exp(\tau/4)$, the VL model with $\beta_{const}(\tau)=\langle \beta_{\Gamma}(\tau) \rangle$, and the SIR model with a single infection rate of $\beta$ respectively. $\beta=\langle \beta_{\Gamma}(\tau) \rangle$ and $\beta = 2m\langle a\rangle \langle \beta_{\Gamma}(\tau) \rangle$ for the configuration and activity models respectively. $R_0=3$ for each infection curve.}
    \label{fig:time_series}
\end{figure}

We simulated all the contagion models described in discrete time with $\Delta t = \Delta \tau = 1$. We simulated the SIR model as a discrete time Markov process using the parameters $\gamma = 1/\tau_R$ and $\beta = \langle \beta_{\Gamma}(\tau)\rangle$ and $\beta =2m \langle a \rangle \langle \beta_{\Gamma}(\tau)\rangle$ for the configuration and activity models respectively. For the viral load model, we store the time at which node $i$ has been infected as $t_i^*$ and at time $t$, the rate of infection of that node is $\beta(t - t_i^*)$, for example, and when $t - t_i^* \geq \tau_R$, the node recovers. When simulating on temporal networks, we store the temporal network as an array, where each entry is a static network corresponding to a particular snapshot in time.

From Fig.~\ref{fig:time_series} we see that the peak of the SIR model is delayed relative to both viral load models and for the static case, the epidemic peak is significantly less pronounced. We comment that the viral load model fundamentally changes the time scale of the epidemic when compared to the SIR model. Not all infectious compartments are created equal, however; someone at their peak infectiousness contributes much more to the spread of an epidemic than someone who has just gotten infected or almost recovered. For this reason, we now plot the number of individuals in each infectious stage over time. We now relax the assumption that $\beta_{\Gamma}(\tau)$ and $\tau_R$ are identical for each member of the population. We assume that $\arg\max_{\tau}\beta(\tau)\sim \text{Uniform}(2, 6)$ and that $\tau_R \sim \text{Uniform}(16,26)$ similar to Ref.~\cite{larremore_test_2020}. At given times $t$, we plot the number of individuals as a function of the infectious duration $\tau_i = t - t_i^*$ and $t$.

We see traveling wave behavior as described in Section~\ref{section:continuum} for both static and temporal networks. The amplitude of this wave varies in response to the introduction of new infected individuals, but the distribution shows a clear transition to the latter infectious stages as the epidemic progresses. This behavior is corroborated by the three normalized vertical cross-sections, showing the probability distribution at selected times. We notice that, despite identical values of $\beta_i(\tau)$ and $\tau_{R_i}$ for every node, the temporal behavior is different for static and temporal networks. For the temporal network, it seems evident that individuals with the longest infection duration seem to be driving the epidemic based on the minimal decrease in individuals for large $\tau$ in comparison to the static network case.

\begin{figure}
    \centering
    \includegraphics[width=8.6cm]{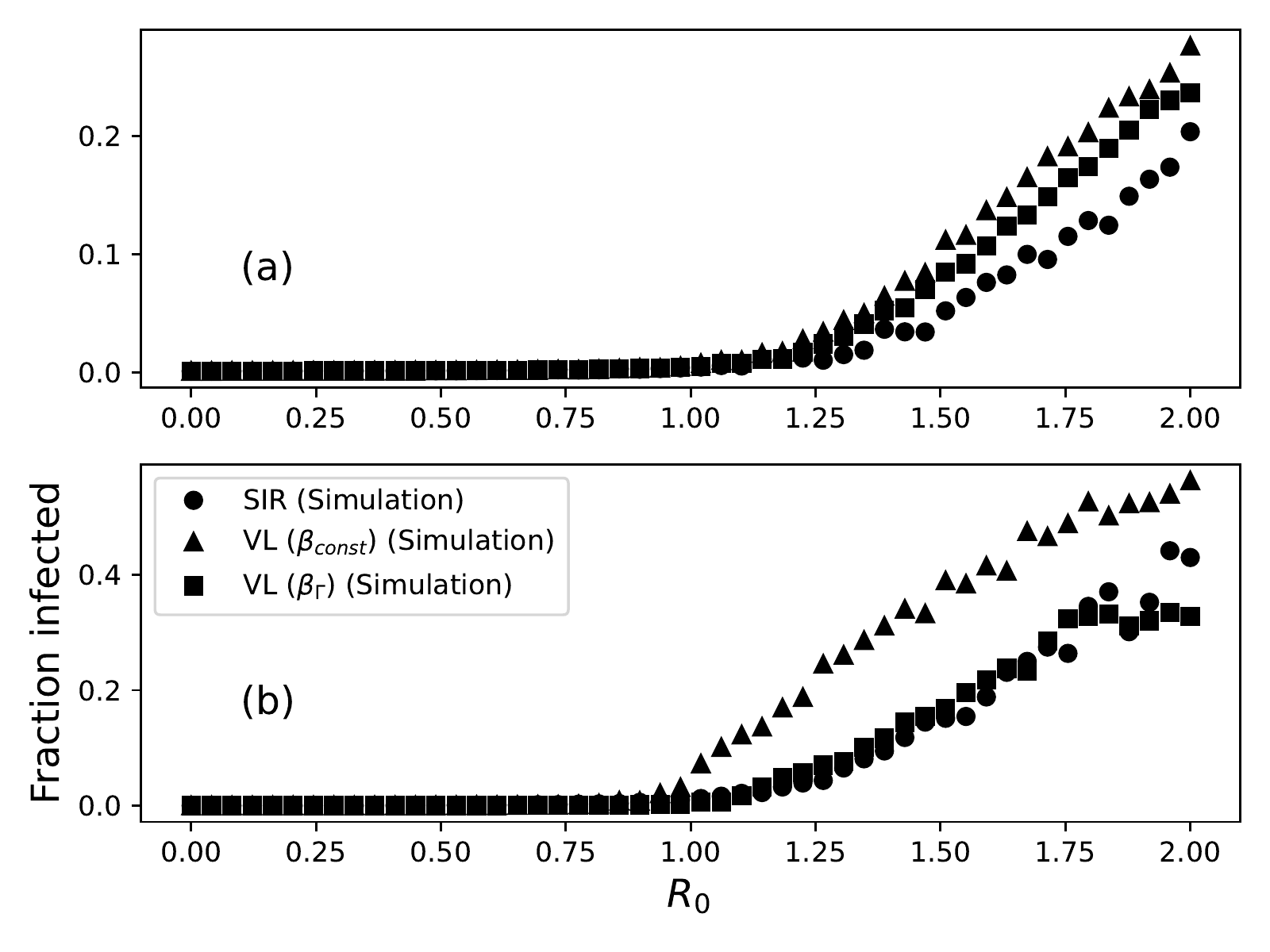}
    \caption{The epidemic extent plotted as a function of the predicted reproductive number for different contagion models for (a) the configuration model and (b) the activity model.}
    \label{fig:extent_vs_R0}
\end{figure}

We also plot the epidemic extent for different values of $R_0$ in Fig.~\ref{fig:extent_vs_R0} to validate our predictions of the reproductive number. For each data point, we averaged over 100 simulations but use the same network realization for all simulations for both the configuration and activity models. We ran each simulation until there were no longer any infected individuals.

We see that for both static and temporal networks, the predictions from our theory do as well as the predictions for the SIR model in Refs.~\cite{boguna_epidemic_2002} and \cite{perra_activity_2012}. The gradual transition is due to the heterogeneity of the networks and agrees with prior results on power-law networks \cite{gou_how_2017}.

\section{\label{section:discussion} Discussion}

In our analysis, we theoretically derived and numerically validated predictions of the population reproductive number for static and temporal networks for a contagion model accounting for time-dependent infectiousness. We see that time-dependent infectiousness causes a fundamental change in the time dynamics compared to the dynamics of the SIR model, despite an epidemic threshold matching classical theory.

Although time-dependent infectiousness does not affect predictions on whether an epidemic will initially grow or die out, it has strong implications how the epidemic progresses in time. In the continuum limit, the viral load model can be written as the transport equation PDE with an infectious boundary condition, which indicates that distribution of $\tau$ progresses in time like a traveling wave and this prediction is validated with numerical simulations.

In this study, we have only considered the population reproductive number, though it is well known that merely studying the population reproductive number without examining the heterogeneity in the number of secondary infections leaves out key information \cite{althouse_superspreading_2020}. Superspreading events are the result of this stochasticity and can often be responsible for the transmission of an epidemic. The VL framework could be used to model the distribution of secondary infections resulting from a combination of contact-based and infectiousness-based heterogeneity.

\section*{Acknowledgements}

I would like to thank Ren Stengel for working on stochastic simulations that were helpful in framing this study, Daniel Larremore for many helpful conversations and theoretical insights, Juan G. Restrepo for helpful suggestions and draft edits, and Subekshya Bidari for her help in formulating the PDE model.

\section*{Data Availability}
All code used in this study can be found at \href{https://github.com/nwlandry/time-dependent-infectiousness}{https://github.com/nwlandry/time-dependent-infectiousness} \cite{landry_time-dependent-infectiousness_2021}.

\bibliography{time_dependent_infectiousness}

\end{document}